%% file: jan6-content-moderation.tex
\documentclass[sigconf,nonacm]{acmart}
\usepackage[english]{babel}
\usepackage[utf8]{inputenc}
\usepackage{booktabs}
\usepackage{multirow}
\usepackage{xcolor}
\usepackage{colortbl}
\usepackage{subcaption}
\usepackage{tabularx}

\title[Understanding the (In)Effectiveness of Content Moderation]{Understanding the (In)Effectiveness of Content Moderation:\\ A Case Study of Facebook in the Context of the U.S.\ Capitol Riot}
\titlenote{Version 1 (January 2023): Non-peer-reviewed technical report.\\
Version 2 (February 2023): Added related work, acknowledgments, and authors.}

\author{Ian Goldstein}
\affiliation{%
 \institution{New York University}
\country{USA}
}

\author{Laura Edelson}
 \orcid{0000-0003-4988-6810}
\affiliation{%
 \institution{New York University}
\country{USA}
}

\author{Minh-Kha Nguyen}
\affiliation{%
 \institution{Université Grenoble Alpes, CNRS, Inria, Grenoble INP}
 \country{France}
}

\author{Oana Goga}
\affiliation{%
   \institution{CNRS, Inria, Institut Polytechnique de Paris}
   \country{France}}

\author{Damon McCoy}
\affiliation{%
 \institution{New York University}
 \country{USA}
}
\author{Tobias Lauinger}
\orcid{0000-0002-5779-0643}
\affiliation{%
 \institution{New York University}
\country{USA}
}

\begin{document}

\begin{abstract}
    \input{sections/0-abstract}
\end{abstract}

\maketitle

\input{sections/1-introduction}
\input{sections/2-background}
\input{sections/3-methodology}
\input{sections/4-results}

\input{sections/5-discussion}
\input{sections/6-conclusion}
\input{sections/7-acknowledgments}
\input{sections/8-ethics}

\bibliographystyle{ACM-Reference-Format}
\bibliography{library}

\end{document}

%% file: sections/0-abstract.tex
Social media networks commonly employ content moderation as a tool to limit the spread of harmful content. 
However, the efficacy of this strategy in limiting the delivery of harmful content to users is not well understood. 
In this paper, we create a framework to quantify the efficacy of content moderation and use our metrics to analyze content removal on Facebook within the U.S.\ news ecosystem.
In a data set of over $2\,M$~posts with $1.6\,B$~user engagements collected from 2,551~U.S.\ news sources
before and during the Capitol Riot on January~6, 2021,
we identify 10,811~removed posts.
We find that the active engagement life cycle of Facebook posts is very short, with 90\,\% of all engagement occurring within the first 30~hours after posting.
Thus, even relatively quick intervention
allowed significant accrual of engagement before removal,
and prevented only 21\,\% of the predicted engagement potential during a baseline period before the U.S.\ Capitol attack.
Nearly a week after the attack,
Facebook began removing older content, but these removals occurred so late in these posts'
engagement life cycles that they disrupted less
than 1\,\% of predicted future engagement, highlighting the limited impact of this intervention.
Content moderation likely has limits in its ability to prevent engagement, especially in a crisis, and we recommend that other approaches such as slowing down the rate of content diffusion be investigated.

%% file: sections/1-introduction.tex
\section{Introduction}

Social media companies such as Google (YouTube) and Meta (Facebook)
have become inextricably linked to contemporary civil society~\cite{GORWA2019} through their communication and content sharing ``platforms.''
Billions of users rely on these services to read or comment on the news, entertain themselves, and conduct business.
Despite stated intentions~\cite{FB_oversight} and the commonly used term ``platform,'' most social media companies
are not impartial channels for users’ communications.
On the one hand, social media companies set the operating parameters of algorithms that promote ``interesting'' content to users.
On the other hand, when it is deemed necessary to retain
users~\cite{Gillespie2018GLR} or to survive public scrutiny~\cite{christchurch_livestream},
social media companies may intervene against extreme or offensive material.

Content moderation
is currently the primary strategy that social media companies employ to counter the spread of harmful content~\cite{FB_moderation,social_media_moderation}.
Nearly all services, including venues for adult (OnlyFans~\cite{onlyfans_terms}) or extreme (8kun~\cite{8kun_guidance}) content, have established policies that dictate what they are willing to host;
content deemed inconsistent with those policies may be removed~\cite{FB_comm_stds}.
Public debate has questioned the adequacy of these policies~\cite{content_moderation_debate} (i.e., \emph{which} content should be deleted) or surfaced instances of objectionable content that was not removed~\cite{Facebook_files}.
Prior research studied potential biases in moderation of YouTube comments~\cite{NEU_YT_2019, NEU_YT_2020}, the impact of content moderation on
user behavior on Reddit~\cite{srinivasan2019content,jhaver2019does}, 
and the magnitude of content moderation regarding third-party links to conspiracy theory stories~\cite{covid19ContentModeration}.
It is still an open question \emph{how quickly} content is removed, independent of the specific policy.
We argue that to judge the effectiveness of content moderation, it is important to understand the spread of content until the time of deletion, i.e., whether content is removed before it reaches a large audience.

In this paper, we
present what we believe to be the first
measurement study
of
the speed and impact of content moderation
on Facebook,
both under normal times and during a crisis.
We were lucky to obtain from Edelson et al.~\cite{misinfo_2021} a data set of public Facebook posts from 2,551~U.S.\ news publishers and ``influencers'' incidentally collected around the time of the U.S.\ Capitol Riot on January~6, 2021.
Because Facebook currently does not make any post-level data transparent about content moderation, we develop a novel methodology to infer content removals and their approximate timing from daily observations of active posts.
For our analysis, we develop novel metrics to estimate the impact of content removal on user engagement, that is, the combined number of comments, shares, and reactions such as ``like.''
We do so both in terms of past engagement that was allowed to occur because of the time it took to remove a post, and in terms of future engagement that was \emph{prevented}.
We estimate the latter based on predictions of the engagement potential of posts, which on aggregate have a net error of 4.5\,\% for viral posts, and 3.2\,\% for normal posts.

For a baseline period selected to represent ``normal'' times, we found that the active engagement life cycle of Facebook posts was very short, with 90\,\% of all engagement occurring within the first 30~hours after posting.
An implication of this is that even relatively quick removal of content (we observed a median of 21~hours) allowed significant accrual of engagement before removal ($3.8\,M$~engagements with 3,843~posts), and prevented only 21.2\,\% of the predicted engagement potential.
By the time a removal decision was made, the content had already reached the vast majority of its typical audience, and the removal had only a
minor impact in terms of preventing engagement or exposure to users.

During the crisis surrounding January~6, we initially observed similar levels of content removals and prevented engagement rates, but with generally higher engagement across both removed and non-removed posts.
It took 6~days, and two announcements of changes to content moderation by Facebook, before we observed a meaningful change in our
metrics.
Our results show that on January~12, Facebook began removing older content, but because these 1,416~removals occurred so late in these posts' engagement life cycles, we estimate that they disrupted less than $1\,\%$ of future engagement, and made hardly any difference in practice.
In this regard, Facebook did not appear to be prepared to contain the fallout of the crisis in a timely manner.

Our results suggest that Facebook's content moderation of U.S.\ news publishers and ``influencers'' could not keep up with the speed at which content spread on their social network, neither during normal times nor in times of crisis.
They highlight that content moderation on a social network such as Facebook operates within the constraints of content recommendation, and its efficacy cannot be assessed in isolation.
Merely tallying numbers of removed posts, without taking into account enforcement delays or the rates at which deleted posts reached their projected audiences, does not adequately characterize the outcomes of content moderation.
Unfortunately, this is not reflected in the transparency metrics that Facebook is currently reporting.

Our work makes the following contributions:
\begin{enumerate}
    \item We propose a methodology for inferring content moderation of public posts from current
    transparency data.
    \item We introduce more insightful metrics for quantifying the impact of delayed content removals on accrued and prevented user engagement.
    \item We show that moderation of public posts by U.S.\ news sources happened late in their engagement life cycle, resulting in only 21.2\,\% prevented engagement.
\end{enumerate}

%% file: sections/2-background.tex
\section{Background \& Related Work}

Facebook and other social media networks reach billions of users by promoting content generated by large content producers such as news organizations or ``influencers,'' and also by individual users.
Unfortunately, social media has been utilized for harmful purposes, such as spreading disinformation~\cite{Benkler2018} or planning and publicizing violent activities~\cite{fagnoni2019terrorism}.
This
has caused these networks to come under increasing societal, regulatory and legal pressures to prevent the promotion of dangerous and harmful content on their systems.

\subsubsection*{Content Moderation Processes.}
Virtually all social media networks have established rules of conduct and acceptable content for their services.
They have a range of techniques at their disposal to enforce these standards, such as deleting~\cite{srinivasan2019content,jhaver2019does}, downranking~\cite{reduction}, quarantining~\cite{quarantining1,quarantining2},
labeling~\cite{warningLabels},
or demonetizing~\cite{demonetization} undesirable content, or temporarily or permanently banning the accounts of users who repeatedly post violating content (``deplatforming'')~\cite{deplatforming1,deplatforming2,FB_trump_ban}.
In this paper, we study \emph{content moderation}~\cite{Gillespie2018,social_media_moderation,FB_moderation,FB_enforcement} from the angle of \emph{removal} of violating content.

The technical and human challenges of implementing reliable mechanisms for content moderation at scale have been topics of repeated study~\cite{commercial_moderation_2019, Helberger2020}.
Social media networks often use automated systems to monitor user content and communications, and ideally interdict violating content before it is published.
Such proactive policy enforcement is typically built as sophisticated pattern matching systems, comparing content to ``blocklists'' of known examples~\cite{Gillespie2020, GORWA2020}, and is thereby limited to predefined classes of violations (graphic violence, sexual content, child abuse, spam)~\cite{moderation_2015}.
To identify new patterns of violations, and address more ambiguous or context-sensitive cases that cannot be handled well by automated systems, social media networks also commonly have content examined by human reviewers after publication, for example when the content reaches a certain popularity threshold, exhibits suspicious interaction patterns, or causes complaints from other users~\cite{flagging,NYT_cleanup_2021}.
Human review of the often violent
content has been documented to have a detrimental effect to the mental well-being of those engaged in the process~\cite{Roberts2016, NYT_cleanup_2021}.

\subsubsection*{Outcomes of Content Moderation.}
Prior research on content moderation (specifically, content removals)
has often focused on legal and socio-political issues (e.g.,~\cite{contentModerationSocial1,contentModerationSocial2,moderation_2015,klonick2017new}), or on qualitative work aiming to understand users' experiences and perceptions of content moderation (e.g.,~\cite{contentModerationInterview1,commercial_moderation_2019,contentModerationInterview2,contentModerationInterview3}).
Quantitative research measuring content removals is more scarce.
Srinivasan et al.~\cite{srinivasan2019content} studied the impact of content removals on subsequent user behavior within a single subreddit, whereas Jhaver et al.~\cite{jhaver2019does} measured the impact that explanations accompanying content removals had on future user behavior across the entire platform.
Jiang et al.~\cite{NEU_YT_2019,NEU_YT_2020} evaluated partisan bias in content and comment moderation on YouTube and found no evidence of left or right bias in removals.
Papakyriakopoulos et al.~\cite{covid19ContentModeration} quantified the sharing of conspiracy theory-related URLs on various social media and modeled the impact of content moderation; in contrast to our work, they did not delve into the timing of engagement accrual or deletion delays.
To the best of our knowledge, we are the first to study the interplay between the timeliness of content removals and the accrual (or prevention) of user
engagement with the removed content.

\subsubsection*{Community Standards on Facebook}
The rules for user behavior and acceptable content on Facebook are called ``Community Standards''~\cite{FB_comm_stds}.
They are divided into six broad
areas ranging from violence and incitement to intellectual property concerns.
These policies have evolved over time, often in response to crises.

Our study period around the U.S.\ Capitol Riot on January~6, 2021 encompasses several such policy changes.
On December~14, 2020, Biden was declared the winner of the Electoral College vote~\cite{election_electoral_college} following the 2020 U.S.\ presidential election that had taken place on November~3.
However, President Trump refused to concede and announced a rally in Washington, D.C.\ for January~6. 
After that ``Stop the Steal'' rally, a group of armed people breached security and stormed the U.S.\ Capitol building~\cite{election_insurrection}. 
Social media platforms were at the center of the spread of false and misleading information about the election, including the use of Facebook to promote and livestream the Capitol attack~\cite{election_misinfo_2, election2021long,jan6_FB_2022}.
In reaction to the attack,
Facebook publicly updated their Community Standards
to explicitly prohibit `praise and support' of the storming of the U.S.\ Capitol, calls to bring weapons to locations in the U.S., video posts and photos from Capitol insurrectionists, violations of the D.C.\ curfew, and future calls to violence.
In addition, Facebook committed to enforcement actions targeting militarized social movements, specifically naming the Oathkeepers and QAnon~\cite{FB_policy_2}.
On January~11, Facebook announced that they would remove content containing ``Stop the Steal''~\cite{FB_policy_1}.
In Section~\ref{sec:analysis:Jan612}, we analyze how these announced policy changes were reflected in observable content moderation activity.

\subsubsection*{Other Related Work.}
More loosely related to our work are prior studies of deleted content (e.g.,~\cite{deletedTweets2013,deletedTweets2016a,deletedTweets2016b}), which however were focused on characterizing users who voluntarily deleted their own content after ``regretting'' its publication rather than measuring content moderation imposed by the social network.
Similarly, there has been ample prior work on the spread of (and engagement with) misinformation on social media (e.g.,~\cite{doi:10.1177/2053168019848554,SHIN2018278,misinfo_2021}), but it has largely left open the question of how fast such misinformation may be removed by the social network.
More generally, researchers have also studied the diffusion of content on social media and the concept of ``viral'' content, for example from the perspectives of user intention~\cite{doi:10.1177/1461444814523726} and the structure of diffusion~\cite{virality2015}.
Studies of viral content
typically take a set number of the `most popular' or widely spreading content.
Examples of this include Pressgrove et al.~\cite{pressgrove2018contagious} with Twitter content, and Vallet et al.~\cite{vallet2015characterizing}
for YouTube content spreading via Twitter.
Past research has also tackled the problem of predicting the popularity of content (e.g., for Facebook and YouTube videos based on visual features~\cite{7903630}).
Unfortunately, our data set does not contain the multimedia content from removed posts, thus we cannot use approaches based on content features to estimate the engagement potential of deleted posts.

%% file: sections/3-methodology.tex
\section{Data Set}
\label{sec:methodology}
After the events of January~6, 2021, we sought to study the changes Facebook made to their content moderation efforts.
Unfortunately, CrowdTangle, Facebook's major transparency tool, does not reveal anything about posts blocked
before publication, and states that once a published post is deleted from
Facebook, it is also removed from CrowdTangle (with a delay).\footnote{https://help.crowdtangle.com/en/articles/3323105-academics-researchers-faq}
Thus, in order to measure post-publication content moderation, it is necessary to
discover newly published posts in near real time, before they may be deleted, but we had not set up a measurement process designed to detect content moderation in advance of the Capitol Riot.
However, we were able to obtain a data set of public Facebook posts by U.S.\ news publishers
collected by the authors of another study~\cite{misinfo_2021}, which happened to include the
time period of interest.
Below, we summarize the original data collection methodology, providing additional
detail where it pertains to the timeliness of detecting new posts.
We then describe our methodology for detecting deleted posts and estimating content
moderation delays in a data set that was not specifically designed for that purpose.
We derive three post data sets for our study, as shown in Table~\ref{tab:dataset_summary}: The 
Removed Set, which we further split into smaller sets for specific analyses; all non-removed
posts from  Facebook pages with at least one deleted post (Impacted Publisher Set); and all non-removed posts across all pages.

\subsection{Monitoring News Publishers on Facebook}
The data set we use in this paper is comprised of the public Facebook posts of 2,551~U.S.\ news publishers during the 2020 U.S.\ presidential election, provided to us by the authors of a study on user engagement with misinformation~\cite{misinfo_2021}.
Their selection of Facebook pages defines the scope of our study.
The authors of that study based their selection of Facebook pages on third-party data provided
by the news rating organizations NewsGuard~\cite{newsguard_2021} and
Media Bias/Fact Check~\cite{mbfc_2021}.
Based on the assessments of these organizations, the authors derived the political leaning of
each news publisher (Far Left to Far Right) as well as a binary ``misinformation'' attribute
indicating whether a news publisher had a known history of spreading
misinformation~\cite{misinfo_2021}.
To exclude largely inactive news publishers, the authors of that study also removed
Facebook pages with fewer than 100~followers or 100~total engagements during their five-month
study period; our study inherits this filter.
Due to the selection of Facebook pages in the original data set, we can study
post-publication moderation of public posts (not comments) on the Facebook pages of U.S.\ news
sources.
These news sources are a mix of publishers known for reliable reporting as well as less
reputable sources with a history of spreading misinformation, both mainstream and niche, from
across the political spectrum.

\subsection{Original Data Collection Methodology}
\label{sec:datacollection}
The original study authors collected all public Facebook posts and corresponding
engagement metadata from the 2,551 U.S.\ news publishers using two types of crawls of the
CrowdTangle API:
(1) A daily crawl to discover all \emph{new posts} since the last crawl, and (2) a second,
separate daily history crawl to update the engagement metadata such as the number of likes,
shares, and top-level comments of \emph{existing posts}.
This latter crawl went back in history as far as the remaining daily time permitted;
in practice, existing posts fell into the observable history window for at least two weeks (or
longer) after publication.
Both crawls were running daily from August~10, 2020 until January~18, 2021.

\subsection{Detecting Deleted Posts \& Post Lifetime}
\label{sec:detecting_deletes}
In September 2021, we performed a follow-up crawl of the same news publisher pages to determine
whether each post from the original data set was still available on CrowdTangle or had been deleted.
This follow-up crawl, however, does not reveal \emph{when} a post was deleted.
To detect deletions with daily granularity, we leverage the daily history crawls.
In the original data set, these history crawls ran until January~18, thus we limit our analysis to
posts published between December~14, 2020 and January~15, 2021 so that we have sufficient buffer to
detect possible deletions.
We conservatively estimate \textbf{post lifetimes} as the time difference between the publication date of the
post (provided by CrowdTangle) and the \emph{last time the post was observed} in a daily history crawl.
This is a lower bound on actual post lifetimes, i.e., we underestimate rather than overestimating post lifetimes, as a post could have been deleted any time in the 24\,h between the
last time the post was observed, and the first time the post was missing in a history crawl.

We do not know why any particular post was removed or whether it was deleted voluntarily or forcibly moderated by Facebook;
we only know that a post is not publicly accessible any more when it fails to show up in the
daily history crawl or our September follow-up crawl.
We also note that very short-lived posts (published and then quickly deleted within less than
24\,h) may not be visible to us, depending on how these events fall between
consecutive
crawls.

\subsection{Data Sets \& Time Periods}
\label{sec:data_slices}
\begin{table*}[t]
    \centering
        \resizebox{\textwidth}{!}{%
                \input{figures/dataset_summary}
                }
    \caption{Overview of data sets. Posts created December~14, 2020--January~15, 2021. Engagement is the total accrued by posts at the time of final observation. We define delayed removals as those occurring more than 30~hours after post creation.
    }
    \label{tab:dataset_summary}
\end{table*}

Table~\ref{tab:dataset_summary} shows a summary of the data sets used in this paper.
Our methodology identified 10,811 posts from 878 U.S.\ news publishers that are no longer accessible via CrowdTangle (i.e., no longer publicly visible on Facebook).
We refer to these posts and associated metadata as the \textbf{removed set}.
We divide the removed set into four subsets.
First, we set aside the subset of posts from
\emph{deleted pages}.
When a page is deleted, the deletion cascades to all of its posts.
Since the deletion times of posts deleted with their page are identical (likely independent from their original
posting times), we
study such pages and their posts separately.
We subdivide the remaining deleted posts into three time periods for more targeted analysis:
\begin{enumerate}
\item The \emph{baseline period} contains posts removed prior to January~4, 2021, and is used to
assess ``normal'' behavior prior to the events around January~6.
\item The \emph{January~6 period} contains posts removed between January~4--11, 2021, and is
used to analyze the immediate reaction in the days leading up to January~6 and just afterwards.
\item The \emph{January~12 period} contains posts removed on January~12 or later.
We analyze this period separately because we noticed increased deletion delays after
January~12 (deleted posts were older than usual); this different deletion behavior may correspond to changes on Facebook's side.
\end{enumerate}

We create two additional data sets of posts that were not deleted.
The \textbf{impacted publisher set} comprises the 297,774 remaining posts from the Facebook
pages that had likely experienced content moderation (at least one post deleted).
This data set serves to investigate the non-moderated content produced by these pages.
To give an overview of the overall ecosystem, the
\textbf{non-removed set} contains all non-deleted posts collected during the observation period. This data set serves as a reference for comparison and consists of 2.76\,M posts from 2,315 news publishers (over 2.5\,B
engagements).

\subsection{Post Engagement and Virality}
\label{sec:viral_methodology}
In the context of Facebook posts, we define \textbf{engagement} as the sum of all types of interactions that users can have with a post (and that are made transparent through CrowdTangle).
Specifically, engagement consists of the number of ``likes'' (and other reaction types available on Facebook, such as ``angry'' or ``sad''), the number of times a post is shared, and the number of top-level comments below the post.
Unfortunately, other types of popularity metrics are not available in CrowdTangle (e.g., the number of times a post has been shown to users,
or
how many unique users have viewed the post), thus we cannot include them in our study.

For posts that were still available in September~2021, we obtain long-term engagement numbers from our
follow-up crawl (Section~\ref{sec:detecting_deletes});
for deleted posts, we use the engagement numbers from the last time they were observed on
CrowdTangle.
Similar to our estimates of post lifetime, engagement numbers for deleted posts are a lower bound.
Additional engagement could have accrued between the last observation and the time the post was removed, but as it is unobserved, it does not factor into our analysis.

To study the speed of engagement accrual of (non-deleted) posts, we leverage time series data extracted from CrowdTangle in September~2021.
These show engagement values in
increasing time steps relative to the publication time of the post.

\subsubsection*{Virality}
In the non-removed data set, posts receive a mean total engagement of $896$ user interactions.
However, the uneven distribution of engagement with content on social media is well known: a small number of posts go viral, and the vast majority do not~\cite{pareto}.
We are not aware of a commonly accepted engagement-based threshold for virality.
We define as \textbf{viral} any post that reaches 14.6\,K engagements (three standard deviations above the mean in the set of all non-removed posts).
This definition of virality is based purely on popularity, independent of the speed of engagement accrual.

\subsection{Estimating Prevented Engagement for Removed Posts} 
\label{sec:methodology_predicted_engagement}
To understand the impact of content removals, it is useful to estimate
how much more user engagement a post might have accrued had it not been
deleted.
We do so by first estimating a post's \textbf{engagement potential}, defined as the total engagement a post would typically receive in the long run (i.e., if it is not deleted).
Because accrual of engagement is different for each post, but potentially
more similar among posts from the same publisher, we estimate a post's engagement potential
based on non-deleted posts from the same Facebook page.
From the engagement time series obtained in September~2021 for the
surviving posts published during the baseline period (which are old
enough that we expect them to have exhausted their engagement potential),
we calculate the mean engagement at each time step over all
non-deleted posts from the same page.
Since our observations show that there is a distinct engagement curve for
viral posts, we estimate engagement separately for viral posts.
For pages with more than 10 non-removed viral posts, we estimate viral
mean engagement per time step individually; for all other pages, we fall
back to using global viral engagement estimates.
We define \textbf{prevented engagement} for a deleted post as the difference between its engagement potential and engagement actually accrued before the deletion.
It is estimated by summing all
engagement on the applicable mean time series after the last observation
of the post.
We likely overestimate prevented engagement because posts may have remained active for up to 24~h after the last
observation in the crawl, thus they might have accrued more real engagement
(and less engagement was prevented) compared to the estimate.
In content moderation, it is desirable to remove policy-violating posts quickly and prevent as much engagement as possible, thus our overestimation of prevented engagement paints content moderation efforts as more effective than they may be.

To validate the quality of these engagement predictions, we tested our
methodology on a sample of 10,000 non-deleted posts published during the
baseline period.
We simulated removals by randomly selecting ``deletion'' timestamps
with selection weights based on the typical lifetimes of deleted posts in
the baseline period (Section~\ref{sec:analysis:impacts}).
The prediction accuracy for engagement potential was 89.7\,\% for non-viral posts, and 84.4\,\% for viral posts.
The net error across all 9,929~non-viral sample post predictions was within 3.2\,\% of the total ground truth engagement potential (lifetime engagement) for these posts (4.5\,\% for the 71~viral posts).
As our intention is to apply our prediction methodology to a large number
of posts and not to draw inferences about the performance of individual
posts, these results suggest that our methodology is sufficiently robust.

%% file: figures/dataset_summary.tex
\begin{tabular}{l@{\hspace{0.2cm}}rrrr}
\toprule
\textbf{Data Set} &   \textbf{Post Count}  &  \textbf{Total Engagement}  &  \textbf{Delayed Removals}  &  \textbf{Delayed Engagement}\\
\midrule
Full Data Set~~\textit{(posts published Dec 14--Jan 15)}  &  2.02\,M & 1.61\,B &   &  \\
\addlinespace
Removed Set~~\textit{(last observation of removed posts)} & 10,811 & 10,889,679 & 1,416 & 2,875,064\\
\hspace{.5cm} Baseline Period~~\textit{(Dec 14--Jan 3)}  & 3,843 & 3,785,464& 281 & 301,741\\
\hspace{.5cm} January 6 Period~~\textit{(Jan 4--Jan 11)}   & 1,171 & 3,302,379 & 60 & 374,066\\
\hspace{.5cm} January 12 Period~~\textit{(Jan 12--Jan 15)}   & 2,300 & 3,257,523 & 1,075 & 2,496,708\\
\hspace{.5cm} Removed Pages~~\textit{(Dec 14--Jan 15)}  & 3,497 & 544,313 & NA & NA\\
\addlinespace
Impacted Publisher Set~~\textit{(posts pub.\ Dec 14--Jan 15)}    &  297,774 & 784,842,977 &  & \\
Non-Removed Set~~\textit{(posts pub.\ Dec 14--Jan 15)}  &2.02\,M &  1.61\,B &  & \\

\bottomrule
\end{tabular}

%% file: sections/4-results.tex
\section{Results}
In this work, we aim to characterize how Facebook moderated public content published on the Facebook pages of U.S.\ news publishers.
When a post is rendered unavailable, we do not know who removed the content.
However, as we discuss more thoroughly in Section~\ref{sec:discussion:moderationVSvoluntary}, the patterns we observe across publishers suggest that the vast majority of post removals were not voluntary, but instances of content moderation, that is, posts deleted by Facebook for violation of their platform policies.
Both user engagement with content and Facebook's apparent content moderation changed in the days during and after the events of January~6, 2021, thus we analyze these time periods separately.

\subsection{Impacts of Content Removals}
\label{sec:analysis:impacts}
To understand Facebook's content moderation performance during ``normal'' times, we begin with a baseline period from December~14, 2020 to January~3, 2021.
During this time, the 2,551~U.S.\ news pages published 1.35\,M~posts.
Facebook users engaged with these posts (i.e., ``liked,'' commented, or shared) a total of 948\,M~times.
In terms of likely content moderation, 3,843~posts (0.28\,\%) were removed from 640 pages during this period, on average 183 removals per day.
However, at the last time they were observed active in our data set, these posts had already accumulated 3.8\,M~user engagements (0.32\,\% of total engagement during the baseline period).
If Facebook deemed these posts inappropriate to remain on their platform, they still allowed a considerable number of users to see and interact with the offending content before reaching the decision to take it down.

\begin{figure}[t]
    \includegraphics[width=1\columnwidth]{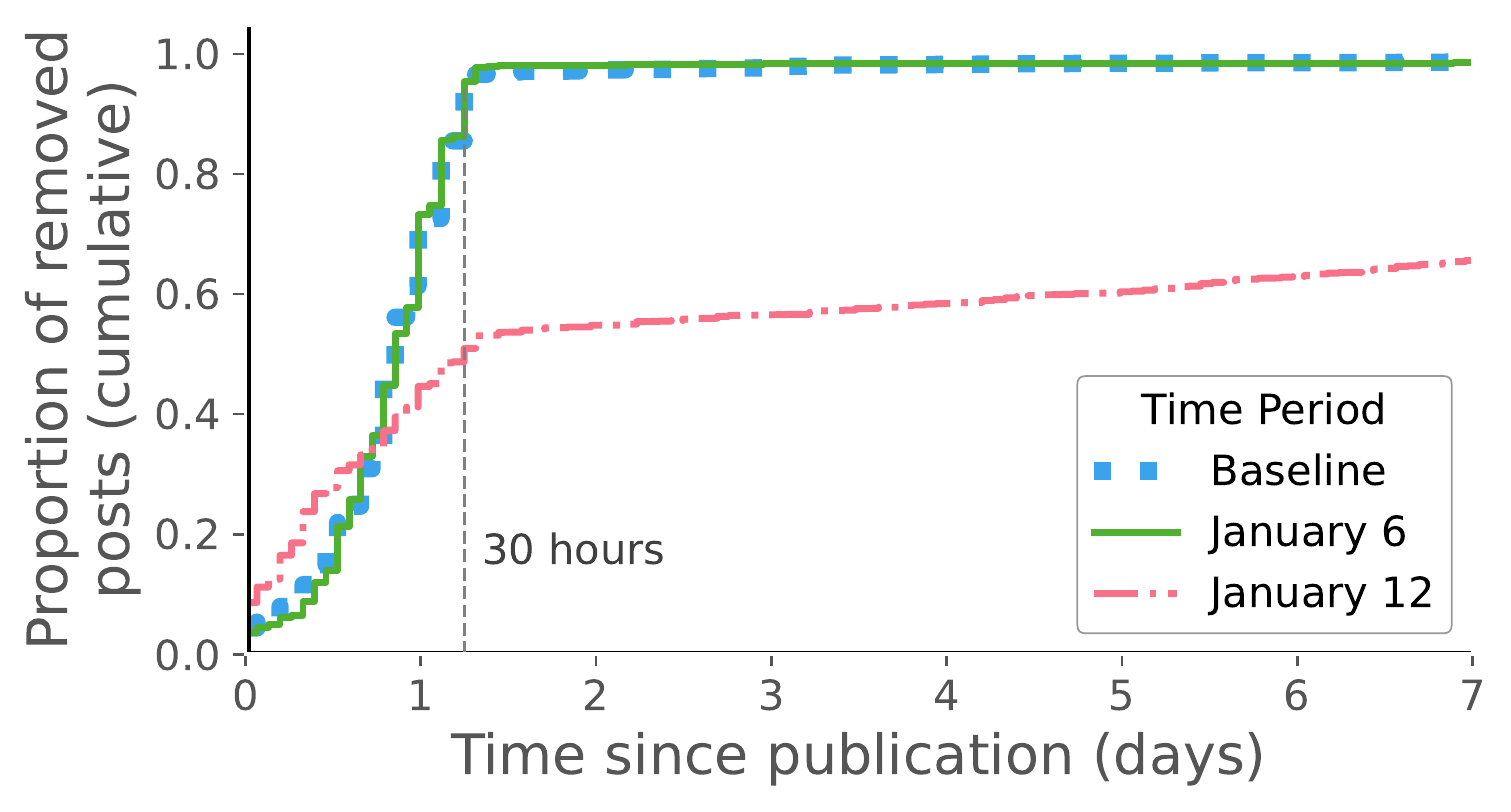}
    \caption{Post lifetime: Days between posting and last observation. In the baseline and January~6 periods, nearly all last observations were at most 30~h after posting, whereas during the January~12 period, much older posts were removed. The sharp pivot at 30~h suggests different removal processes.}
    \label{fig:post_age}
\end{figure}

We turn to the lifetimes of removed posts to understand how these presumably policy-violating posts could accrue so much engagement before being deleted.
As Figure~\ref{fig:post_age} shows, the average time between publication of a post to its last observation by the crawler is 23.7~hours (median: 21~hours).
(Note that the actual time of removal could be up to 24~hours later because the crawler checked a post's status only once every 24~hours, thus we likely underestimate post lifetimes.)
The distribution of post lifetimes pivots 30~hours after post creation; approximately 90\,\% of removed posts have had their last observation at this point.
The remaining 10\,\% are deleted at a noticeably slower pace over a much longer time span, which suggests that a different process might be at play for those deletions.
In summary, most decisions to take down posts appear to be relatively fast in absolute terms, but they do take time, and significant amounts of engagement are allowed to accrue during that time.

The situation of viral posts is slightly different from the overall post distribution.
The baseline period saw 9,768~posts accumulate sufficient engagement to be considered viral (425\,M combined); 61 of them were removed during this period (more than twice the rate of all content removals).
Until their last observation, these 61 removed viral posts had been able to accrue 1.7\,M engagements, 44.8\,\% of all engagement with posts removed during this period.
This share illustrates the importance of handling viral posts well; quick intervention on a relatively small number of violating viral posts could disproportionately reduce the odds of Facebook users seeing and engaging with inappropriate content.
However, our data suggest that viral posts may be more \emph{difficult} for Facebook to handle, as viral posts tended to be active for longer before being removed (mean time to last observation: 31~hours, more than 7~hours longer than the overall mean; median: 24~hours, or 3~hours longer).

\begin{figure}[t]
    \includegraphics[width=1\columnwidth]{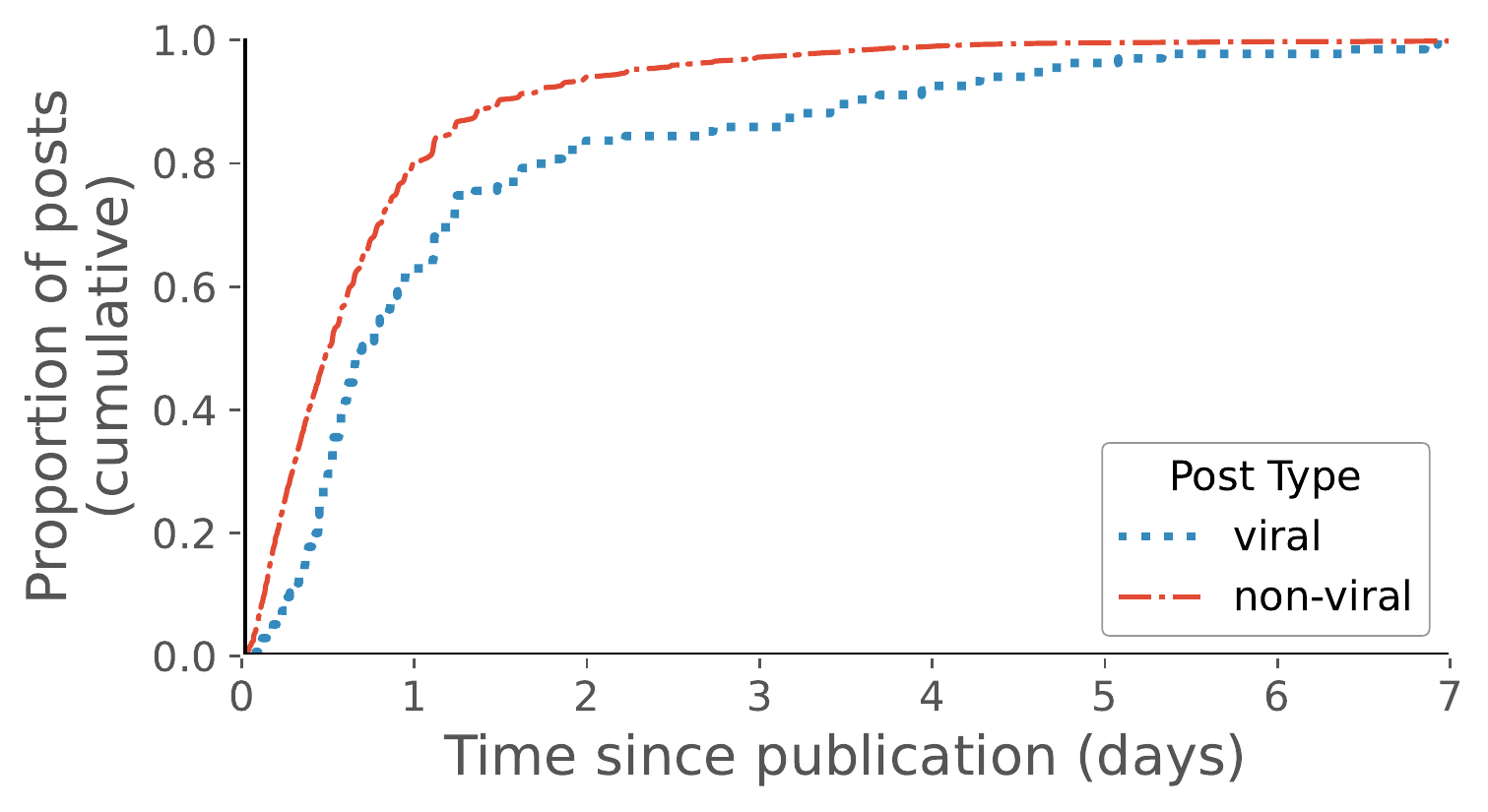}
    \caption{Days until non-removed baseline posts reach 80\,\% of their lifetime engagement. Around 78\,\% of non-viral posts reach 80\,\% of their lifetime engagement within one day, whereas it takes two days for viral posts.}
    \label{fig:time_til_80}
\end{figure}

To put the speed of deletions into context, we compare it to the rate of accrual of engagement.
Most engagement with posts in our data set happens relatively soon after the content is created. 
This is true both for viral and non-viral posts, as shown in Figure~\ref{fig:time_til_80}.
Among non-removed content created during our baseline period, 78.1\,\% of normal posts reached at least 80\,\% of their total engagement within one day, while 93.3\,\% of normal posts achieved it within two days.
In other words, while the typical content moderation delays observed in our data set may seem fast in absolute terms, in fact they do come late in a post's life cycle because most of a post's engagement usually occurs \emph{before} the typical content moderation delays.
As a result, these instances of content moderation may not make a big difference in practice because most of a post's audience will already have seen and interacted with the post before it is deleted.

Viral posts have a longer period of active engagement accrual compared to non-viral posts, with 55.2\,\% reaching 80\,\% of their total engagement in one day, and 80.6\,\% reaching it within two days.
However, this does not mean that slower content moderation would be appropriate due to the slower relative rate of engagement accrual.
In fact, total engagement with viral posts is so much higher that despite the slower relative increase,
the absolute increase is substantial.
In the hour before their respective median removal time, viral posts accrued an average of 1,583~additional engagements, whereas it was only 7.5 for non-viral posts.
With the goal of reducing total exposure to harmful content, it is important to moderate viral posts with the same (or higher) speed than non-viral posts, yet our data suggest the opposite is occurring in practice.

We assess the benefit of typical content moderation delays by estimating how much engagement was prevented by removing the post, that is, how much of the typical audience did \emph{not} see and interact with the offending post.
We estimate a post's typical engagement on a per-page basis and separately for viral and non-viral posts, as described in Section \ref{sec:methodology_predicted_engagement}, and calculate prevented engagement as the difference between that estimate and the last observed engagement count before deletion.
During the baseline period, we estimate that post removals disrupted 21.2\,\% of potential engagement (314 or fewer prevented engagements each for 90\,\% of removed posts).
For the 3,782 removed non-viral posts, this breaks down to a 20.5\,\% prevention rate, or a mean of 151 prevented engagements per non-viral post.
The removal of each of the 61~viral posts had a much larger impact in absolute terms, with a mean of 7,289 prevented engagements.
Viral posts accrue engagement for a longer time (median 1.4 times longer) than non-viral posts, offering the opportunity to disrupt a larger fraction of their engagement potential, but because they were also removed more slowly (median 1.2 times slower), the engagement prevention rate for viral posts was only slightly higher at 22.0\,\%.
Timelier enforcement of the comparatively few removed viral posts could have made an outsize difference in preventing Facebook users from engaging with subsequently moderated posts.
Overall, we estimate that the content removals we observed during the baseline period prevented 996\,K engagements from occurring, which is a consequential number in absolute terms.
Unfortunately though, these removals came so late that the posts had already reached over three quarters of their engagement potential, and only a small minority of their potential engagement was actually prevented by the removal.

\subsection{Content Removals After the Capitol Riot}
\label{sec:analysis:Jan612}
To understand how Facebook carried out content moderation during and after the Capitol Riot, we repeat the analyses of the baseline period above for the time just before and after the event of January~6, 2021.
In the eight days from January~4--11, which we refer to as the January~6 period, we observed comparable levels of content removals in absolute terms, but higher engagement with content by the Facebook users.
In detail, the U.S.\ news pages published 393\,K~posts that received a total of 409\,M~engagements.
The period saw 1,171~posts removed from 246 pages, a mean of 146 per day, which is roughly comparable to the baseline period ($t=1.83$, $p=0.079$).
However, by the time they were last observed, these posts had received 3.3\,M~engagements, 0.81\,\% of all user engagement,
significantly ($t=4.06$, $p<0.001$) more than during the baseline period.
The median post lifetime in the January~6 period was consistent with the baseline period for non-viral ($\chi^2=1.24$, $p=0.266$) and viral posts ($\chi^2=.043$, $p=0.835$), also evidenced by the closely overlapping curves seen in Figure~\ref{fig:post_age}. Thus, the increased engagement accrual did not stem from slower removals, but higher user engagement.
Although we saw no increase in the proportion of posts moderated during the January~6 period, and no significant increase of non-removed viral posts ($t=0.62$, $p=0.536$), there was a significant increase in the removal of viral posts as compared with the baseline period ($t=2.38$, $p=0.024$), which explains the increase in engagement-weighted removals.
80\% of engagement with removed viral posts in this period was associated with extremely partisan sources, and the remainder originated from a single page. 
We estimate that post removals during the January~6 period prevented a similar share of potential engagement (25.8\,\%) as in the baseline period (a total of 306\,K engagements with 1,117 non-viral posts, and 521\,K engagements with 54 viral posts).
In summary, while we observed increased engagement from users, any changes Facebook may have made to content moderation in the immediate aftermath of January~6 do not stand out in our
metrics.

We do observe increased levels of content removals during the four days from January~12--15 (the January~12 period), after Facebook announced another content moderation policy change~\cite{FB_policy_2} on January 11th.
U.S.\ news pages published 186\,K posts during these four days, reaching 161\,M total engagements.
Facebook removed 12~entire pages with all their 3,463~posts created between December~14, 2020 and January~15, 2021.
We discuss these removed pages in Section~\ref{sec:page_deletions}.
In addition, we observed 2,300~removals of individual posts from 377 pages, a mean of 575 per day, which was a significant increase ($t=8.63$, $p<0.001$) from the baseline period.
These posts had received a total of 3.3\,M engagements at the time of their final observation.
This last observation came considerably later in a post's lifetime than in the baseline and January~6 periods.
In the January~12 period, the mean time between post creation and last observation was 152~hours (median: 24~hours), significantly longer than the baseline period for both non-viral ($\chi^2=126.7$, $p<0.001$) and viral posts ($\chi^2=36.8$, $p<0.001$).
As shown in Figure~\ref{fig:post_age}, 48\,\% of last observations occurred more than 30~hours after publication, whereas it was only approximately 6.8\,\% in the baseline and January~6 periods.
This sharp increase in delayed content removals suggests that Facebook may have applied changes in content moderation announced on January~11 retroactively to older posts, which were originally published in the time around January~6, but deleted only after January~11.
(We note that all of the posts removed after more than 30~hours were from ``repeat offender'' pages, i.e., we had already observed removed posts from these pages in the past.)
This further suggests that a policy change, which took nearly a week to implement, was necessary for Facebook to deal with the fallout of January~6 on their platform.

The unusually late removals are reflected in our estimates of prevented engagement.
For all posts removed during the January~12 period, we estimate that 569\,K~engagements were disrupted, representing only 14.9\,\% of removed posts' predicted engagement potential.
In particular, for the subset of posts removed with a long delay (of more than 30~hours), the deletions appear almost inconsequential because they prevented only 0.60\,\% of the engagement we predicted these posts to achieve without being deleted.
In other words, if these deletions were indeed instances of ``retroactive'' content moderation, Facebook allowed these posts to reach virtually their entire potential audience before adjudicating that they were in violation of platform policy and had to be taken down.

\subsection{Impact of Removals by Misinformation and Partisanship Reputation}
\label{sec:removal_misinfo}
Over the last several years, there has been intense public scrutiny of the partisan impact of content moderation policies on Facebook~\cite{gallo2021social, pew_censorship, cato_censorship}.
We now analyze content removals taking into account the partisanship and factualness classifications in our data set.
These labels reflect the reputation of the pages in question rather than characterizing individual posts,
but they still allow us to compare the effects of content removal along partisan lines.
We calculate the engagement-weighted \textbf{rate of removal} as the observed engagement from removed posts in a category of pages over the total observed engagement from all posts in that category.

Over the full time range of our data set, posts made by pages with a reputation for misinformation had a higher engagement-weighted rate of removal
than posts from pages not classified as misinformation providers.
This finding of higher misinformation removal rates held true within each of the five partisanship categories,
as shown in Table \ref{tab:removal_rates}.
At the same time, within the same factualness category, engagement-weighted rates of removal varied considerably according to the partisanship of pages; in the misinformation category, they ranged from 1.02\,\% for posts made by Far Left pages to 6.97\,\% for posts from Slightly Right pages.
We note that misinformation pages of extreme partisanship (Far Left and Far Right) had the \emph{lowest} post removal rates among misinformation pages, whereas non-misinformation pages of extreme partisanship had the \emph{highest} post removal rates of all
non-misinformation pages.
We do not currently have an explanation for this effect.

\begin{table}[t]
        \centering
        \input{figures/removal_rate_table}
    \caption{Post removal rates by partisanship and factualness of the source (weighted by engagement, entire data set). Across the political spectrum, news sources known to spread misinformation saw a higher fraction of their accumulated
    user engagement affected by content removals.}
    \label{tab:removal_rates}
\end{table}

\begin{figure}[t]
    \includegraphics[trim=1.3cm 0 0.9cm 0,clip,width=\columnwidth]{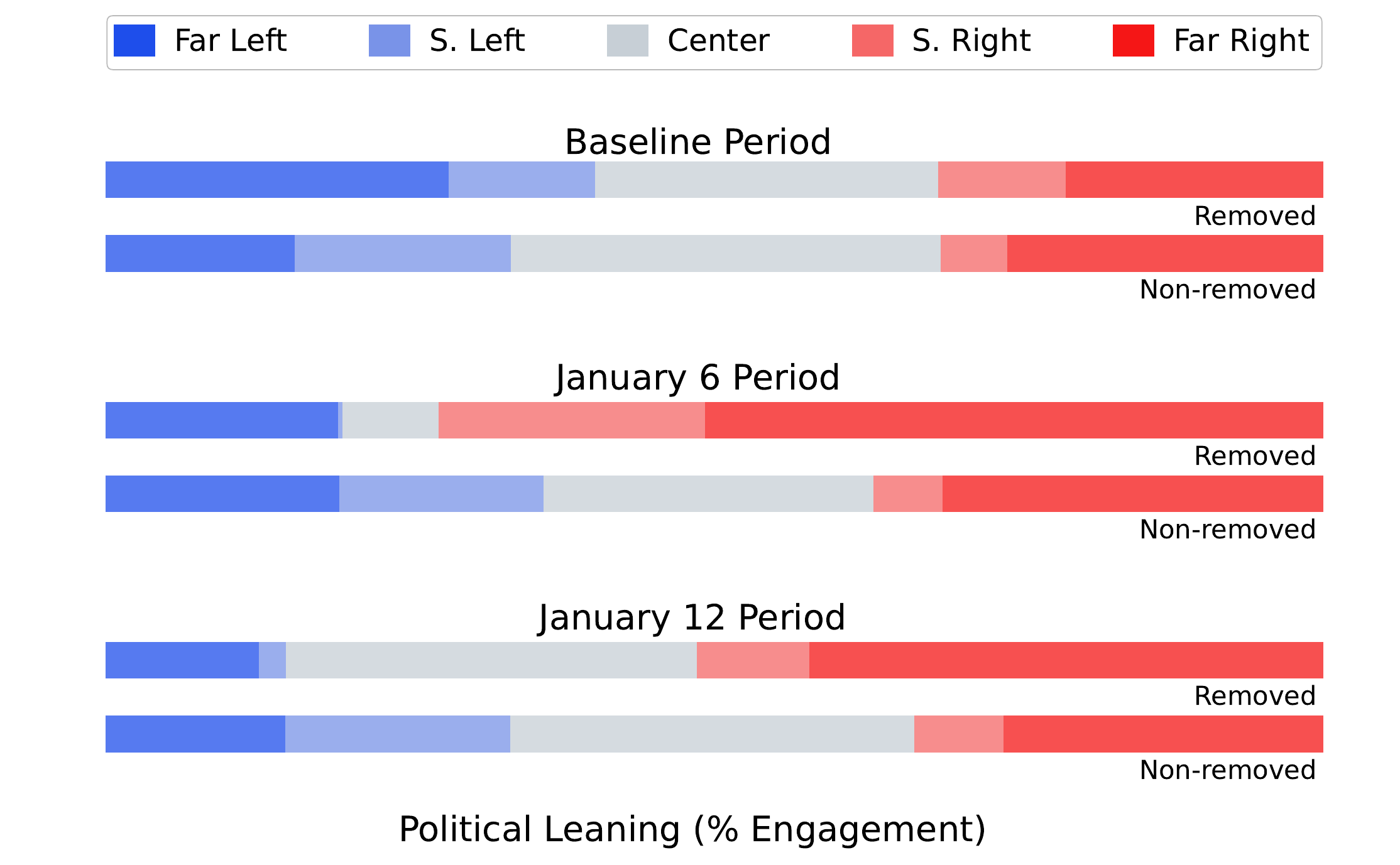}
    \caption{Engagement with removed and non-removed posts from U.S.\ news pages by partisanship
    for the baseline, January~6, and January~12 periods, respectively. Around January~6, the vast majority of engagement with posts that were ultimately removed went to posts from sources classified as Slightly Right and Far Right. The rate of engagement with misinformation sources across all time periods was 66.1\,\% for removed content and 28.2\,\% for non-removed content.}
    \label{fig:eng_dist}
\end{figure}

So far, our analysis of removal rates covered the entire duration of our data set, about a month roughly split in half by the events of January~6, 2021.
These events were of a deeply partisan nature; we now investigate whether any corresponding effects based on partisanship are observable among the removed posts.
To better isolate the impact of these events, we again split the data set into the baseline, January~6, and January~12 periods.
(To streamline the analysis, we no longer differentiate pages by their misinformation reputation.)
The upper two rows of Figure~\ref{fig:eng_dist} show the engagement-weighted partisan breakdown of posts removed during the baseline period (above), and non-removed posts published during the same period (below).
While not perfectly balanced, there does not appear to be any major partisan bias in the posts removed during the baseline period compared to the posts that were not removed.

The picture looks very different in the January~6 period (middle two rows of Figure~\ref{fig:eng_dist}).
The majority of engagement with posts removed during these eight days corresponded to Far Right pages (50.8\,\%), while Slightly Right pages accounted for the next largest share (21.9\,\%).
The partisan breakdown of engagement with posts that were not removed does not exhibit this effect, and more closely resembles the baseline period (albeit not exactly).
During the four days of the January~12 period, the proportion of engagement attributable to removed posts from Slightly Right pages receded to a level comparable to the baseline period.
However, the corresponding proportion from Far Right pages remained more than double the baseline share of engagement-weighted post removals.
(This does not necessarily correspond to ``new'' engagement accrued during this period, but may be partially due to delayed enforcement; 65.7\,\% of these removals were posts originally published in the days around January~6, and deleted only on or after January~12.)
In conclusion, the partisan breakdown of engagement-weighted removal rates was somewhat balanced along partisan lines before the events of January~6, and remained at comparable levels across all three time periods for non-removed posts.
Likely content moderation around January~6 disproportionately affected posts from Slightly Right and Far Right pages.
This is in line with the policy changes that Facebook announced on January~6 and January~11, which specifically targeted so-called ``Stop the Steal''~\cite{FB_policy_1,FB_policy_2} misinformation.

\subsection{Most Impacted Publishers}
For illustrative purposes to complement our quantitative findings, we identified the U.S.\ news pages with most user engagement that likely experienced content moderation during our entire observation period (the Impacted Publisher Set).
The two most popular pages with significant post removals were `Dan Bongino' and `Occupy Democrats,'
both classified as sources of misinformation with Far Right\footnote{https://mediabiasfactcheck.com/dan-bongino/} and Far Left\footnote{https://mediabiasfactcheck.com/occupy-democrats/} partisanship, respectively.
Taken together, both pages generated 835\,K engagements with content that was later removed, though this represented only a fraction of the tens of millions of engagements each of these pages generated in our observation window.
`The Blaze,' a conservative\footnote{https://mediabiasfactcheck.com/the-blaze/} media company with a reputation for misinformation, received the greatest total engagement (1.37\,M) with 310~ultimately removed posts, accounting for 28.3\,\% of all engagement that the page received between December~14, 2020 and January~15, 2021. 
At 91.5\,\%, the share of engagement attributed to deleted posts was highest for `I Love My Freedom,' a page associated with a conservatively oriented online apparel retailer.\footnote{https://www.facebook.com/ILMFOrg/about} 
Three quarters
of engagement-weighted removals from this page occurred after Facebook's January~11 policy announcement, with a mean delay of 8.5~days, suggesting that the affected content might not have been in clear violation of the policies in effect prior to that date.
In line with the unusually long removal delays, we estimate that only 6.2\,\% of potential engagement was prevented.
In summary, the pages most heavily impacted by removed content tended to be pages
categorized as extremely partisan, misinformation providers, or both.

\subsection{Page Deletions}
\label{sec:page_deletions}
So far, we have only considered removals of individual posts from pages that themselves remained active.
In addition,
we observed the removal of 14~entire pages (the Removed Pages Set), which included their 3,497~posts created between December~14, 2020 and January~15, 2021.
These posts had accumulated an engagement of 544\,K.
Deleted pages fall into three categories: known sources of health and vaccine misinformation (`Erin at Health Nut News,' `LifeSiteNews.com'), extreme Far Right media with poor records for factualness and newsworthiness (`TruNews,' `ConservativeOpinion.com'), and pages frequently linked to conspiracy sources (`The Drudge Report,' `Zero Hedge'). 
Many of these were widely known as spreaders of inaccurate information, and their forced deplatforming by Facebook and the related debate were widely reported~\cite{viceDisinfo,VFzerohedge,APzerohedge}.

%% file: figures/removal_rate_table.tex
\definecolor{farLeft}{HTML}{0A337F}
\definecolor{slightlyLeft}{HTML}{3979EF}
\definecolor{center}{HTML}{999999}
\definecolor{slightlyRight}{HTML}{FE1601}
\definecolor{farRight}{HTML}{A70F01}
\definecolor{misinfoBackground}{HTML}{EEEEEE}
\definecolor{nonMisinfoBackground}{HTML}{FFFFFF}
\newcolumntype{a}{>{\columncolor{misinfoBackground}}r}
\begin{tabular}{lrr}
\toprule
\rowcolor{white}
  \textbf{Partisanship} & \textbf{Misinformation} & \textbf{Non-misinformation} \\
\midrule
{\textcolor{farLeft}{Far Left}} &                  1.02\,\% &                      0.57\,\% \\
{\textcolor{slightlyLeft}{Slightly Left}} &                  2.67\,\% &                      0.19\,\% \\
{\textcolor{center}{Center}} &                  3.34\,\% &                      0.29\,\% \\
{\textcolor{slightlyRight}{Slightly Right}} &                  6.97\,\% &                      0.07\,\% \\
{\textcolor{farRight}{Far Right}} &                  1.19\,\% &                      0.53\,\% \\
\bottomrule
\end{tabular}

%% file: sections/5-discussion.tex
\section{Discussion}
Content moderation aims to limit the exposure of users to harmful content by way of deleting user-generated content that is deemed unacceptable.
In this framework, content moderation needs to compete with content recommendation systems, in that the speed of enforcement needs to keep up with the potentially viral spread of toxic content in order to effectively contain undesirable content.

Our results suggest that Facebook's content moderation was not effective at meaningfully limiting the spread of harmful content from U.S.\ news publishers.
In the most common case, content moderation was relatively quick in absolute terms; half of removed posts in the baseline period were last observed 21~hours after publication.
Yet, the speed of removal pales in comparison to how quickly content accrues engagement, and how quickly interest in content fades.
By the time of removal, the posts in our data set had already been shown to millions of Facebook users, accumulating a total of 10.9~million engagements.
We estimate that the removals prevented only 21.2\,\% of engagement from occurring because the vast majority of the posts' potential audience had already seen them.
Furthermore, it appears that Facebook needed to change their policy in order to moderate some content after the Capitol riot of January~6, 2021.
This policy change took nearly a week to implement, likely delaying the removal of 995~posts by multiple days, with the result that they received over 2\,M engagements in the meantime, and their removal prevented virtually none ($<1\,\%$).

Whether or not content moderation can be effective depends on factors that are under Facebook's control.
For one, Facebook could aim to delete content more quickly by decreasing delays in human review and implementing a content policy that does not require lengthy adjustments to allow reaction to major real-world events.
It is unclear, however, how much potential there is for such improvements.
As we have shown, the currently very short active engagement periods of posts, and the speed of viral accrual of engagement create a limited window during which content removal can be effective as a strategy.
Therefore, the second avenue to making content moderation more effective would be changing the parameters of the content recommendation algorithm to enlarge this time window and give Facebook more time to intervene effectively.
This could take various forms.
To increase the time for intervention, new content could be recommended more slowly.
To limit the impact of harmful content, the relatively small number of
posts predicted to ``go viral'' could be reviewed manually before they accrue significant engagement.

Admittedly, the scenario for content moderation that we studied in this paper may be more challenging than the ecosystem of Facebook as a whole.
Public posts from U.S.\ news publishers, due to their newsworthiness, may require a more thorough review than other types of policy violations.
Furthermore, due to their larger audience, delayed removals have a much larger aggregate impact than moderation of posts in smaller groups, or private messages.
These challenges highlight, however, the limits of what can be achieved with the tool of content moderation under the constraints of a recommendation system that has been designed and optimized to spread engaging content as quickly as possible.

\subsection{Limitations}
\label{sec:discussion_limitations}
At the time of this writing, Facebook does not provide
fine-grained transparency for content moderation.
Our efforts to study content moderation rely on
what is observable indirectly through CrowdTangle.
As a result, we can only study posts deleted after
publication.
Other types of content moderation, such as rejecting posts before publication, or deliberately downranking posts in the recommendation algorithm (without deleting them) are outside the scope of our study.
Facebook claims that up to 90\,\% of content violations (including hate speech, nudity, violence, and bullying) are deleted before they are likely to have been seen \cite{FB_enforcement}.
Our study can only include posts that slip through
pre-publication content filters.
Yet, collectively Facebook users engaged with the
deleted posts in our data set more than 10.8~million
times.
While we do not have exact numbers about how this relates to impressions and unique users, it is reasonable to estimate that millions of Facebook users were exposed to these posts that Facebook later deemed inappropriate.

Because we rely on a repurposed data set that was not
originally collected to study content moderation, our
analysis is subject to additional limitations.
The authors of the data set did not archive multimedia
content such as images or videos associated with posts.
Posts may have been deleted for offending image
or video content, thus we cannot conduct a meaningful
study of the contents of deleted posts.
Furthermore, the timing of post discovery in the original
crawl causes a bias towards longer-lived posts, that is,
when posts are deleted quickly after publication, they
are more likely to be missing in the data set.
Similarly, the daily status checks of older posts in the
original crawl mean that the inferred post lifetimes
are only a rough estimate.

Our analysis is based on public posts published by the
Facebook pages of 2,551 U.S.\ news organizations.
The content produced by these pages is not representative
of the Facebook ecosystem as a whole.
Furthermore, the selection of these pages and
their categorization with regard to their
political leaning and history of spreading misinformation is subject to the methodology
and limitations described by Edelson et al.~\cite{misinfo_2021}.

\subsection{Content Moderation vs.\ Voluntary Removals}
\label{sec:discussion:moderationVSvoluntary}
Because we cannot observe post deletions directly, we do not know for certain why content was removed;
it could have been moderated by Facebook, deleted by the
page owner, or the visibility could have been changed to
private.
However, taken together, several independent observations from our analysis suggest that a majority of removed posts were indeed subject to content moderation by Facebook.
First, the pattern of deleted post lifetimes with a sharp pivot at 30~hours was consistent across pages, which suggests intervention by a common process as opposed to independent voluntary deletions by page owners.
We were able to confirm one instance of voluntary deletions by contacting the page administrators of the National Center for Missing \& Exploited Children,\footnote{https://www.missingkids.org/} a non-partisan, charitable organization who confirmed to us that they routinely take down posts about missing children when those children are found.
Statistically, their 185~voluntary removals in our data set (with a total of 344\,K engagements) looked very different from aggregate removal patterns, suggesting that the bulk of removals are not voluntary.
In terms of post lifetimes, we last observed their voluntarily deleted posts after a mean of 7.3~days (median: 2.4~days), as opposed to 2.7~days (0.9~days) across all removed posts in our data set.
This supports our hypothesis based on patterns in post lifetimes that most observed removals were likely deletions by Facebook.
Second, the unusually long removal delays only observed in the January~12 period and the coinciding, prior announcement by Facebook of a content moderation policy change suggest that these removals were instances of content moderation.
Third, the higher prevalence of removals among pages with a reputation for spreading misinformation makes it less likely these are voluntary deletions, unless we hypothesize that these misinformation providers are more prone to self-censorship.
While only improved transparency from Facebook could provide clarity, we are confident that the removal patterns we observed are predominantly a reflection of content moderation.

\subsection{Improved Measurement Design}
If we were to design a measurement of content moderation of public posts from scratch, and assuming
that a transparency tool such as CrowdTangle continues to impose low rate limits and does not
provide specific transparency around content moderation, what would our measurement design look like?

\begin{itemize}
\item \emph{Archive multimedia content} such as images and video.
If space is an issue, keep only the content of deleted posts, or of all posts from pages that have
deleted posts.

\item \emph{Detect new posts more frequently} (as opposed to daily) to observe short-lived
posts.
Prioritize the time of day when most posts are published.

\item \emph{Check the status of existing posts more frequently} than daily to get more
fine-grained post lifetimes, at least during the first 2-3 days, which see the vast majority of
deletions.
\end{itemize}

\noindent For example, hourly daytime crawls to detect publication or deletion of recent posts (0-3 days old)
could be combined with nightly history
crawls to detect deletion of older posts with daily granularity.

\subsection{Transparency Metrics \& Reporting}
Meta's primary mechanism for making data relating to content moderation transparent is their Community Standards Enforcement Report~\cite{FB_enforcement}.
In these quarterly reports, the company shares aggregate, global data about eleven categories of removals,
reported as the \emph{absolute} number of posts removed, and the \emph{relative} share of global user impressions attributable to moderated content.
The discrepancy between absolute and relative numbers makes them difficult to compare, and they are not broken down at a country level.
Furthermore, the raw number of removed posts is not \emph{per se} informative for understanding the efficacy of content moderation.
(A more detailed discussion of Facebook content moderation transparency metrics can be found in the 2019 report of the Facebook Data Transparency Advisory Group~\cite{Bradford:2019vv}.)
Based on the metrics we used in our analysis, we recommend reporting how long content remains active before it is taken down, how many people are exposed to content that is later removed,
and how much of a removed post's potential audience did \emph{not} see the post as a result of the intervention, that is, whether removal of the post made a meaningful difference in safeguarding users.

%% file: sections/6-conclusion.tex
\section{Conclusion}
In this work, we repurposed a data set of over 2\,M Facebook posts from 2,551~U.S.\ news sources
In the absence of explicit transparency about content moderation, we developed a
methodology to identify removed posts within an existing data set, and infer when removals occurred.
We also developed a method for predicting lifetime engagement of posts based on the engagement of other posts from the same Facebook page, and proposed novel metrics to quantify the impact of content moderation based on accrued and prevented user engagement.
We found that content moderation during ``normal'' times was relatively quick in absolute terms (median post lifetime of 21~hours), yet posts tended to exhaust their engagement potential at an even faster rate, and we estimate that at most 21.3\,\% of predicted future engagement was prevented over the entire data set.
In summary
moderation of public content from U.S.\ news sources and influencers on Facebook was too slow to keep up with the spread of content and ensuing engagement;
it is unclear whether this race to limit user exposure to harmful content can be won purely by speeding up content moderation.
We recommend that other strategies be explored further, such as changes to content recommendation algorithms
that slow the diffusion of content through social networks overall.

%% file: sections/7-acknowledgments.tex
\begin{acks}
This research was supported in part by the Democracy Fund, the French National Research Agency (ANR) under ANR-17-CE23-0014, ANR-21-CE23-0031-02, by the MIAI@Grenoble Alpes ANR-19-P3IA-0003, and by the EU 101041223, 101021377 and 952215 grants.
\end{acks}

%% file: sections/8-ethics.tex
\section{Ethics Statement}

In conducting this research and creating this paper, we honored the guidelines and principles established for the conduct of ethical research. 
We have laid out the limitations of our data and methods in
Section~\ref{sec:discussion_limitations}.

In our research, we relied on an existing data set created using
CrowdTangle, an official tool from Facebook, with API access granted to
both us and the original authors~\cite{misinfo_2021}.
CrowdTangle contains only public posts of Facebook pages and aggregate,
purely quantitative engagement data, but no personally identifiable
information.
We did not have access to information about individual news consumers, and
expose no personally identifiable information.

The purpose of our paper is to measure the circumstances and
impact of content moderation on social media, particularly
during a
crisis.
We are cognizant that
our work could cast a poor light on Facebook, or on the
authors of
moderated posts, who might object to a work that emphasizes the
classification of some of their content as objectionable.  
We argue that the benefit of better understanding content moderation
outweighs the potential impact to reputation, particularly when Facebook's conduct has been thoroughly reported~\cite{election_misinfo_1, election_misinfo_2}. 

In identifying the strong connection between removed content and sources'
reputations for mis- and disinformation (Section~\ref{sec:removal_misinfo}), our
intention is not to adjudge the merits of any publisher, merely to record
engagement based on existing reputations for factualness, for which we
rely on multiple third-party sources for classification.
Ultimately, we reason that it is a better outcome for society in general
that Facebook not become a haven for harmful content than that any
of these stakeholders maintain a sterling reputation.
To this end, we contribute
recommendations for improving
moderation transparency and efficacy for the benefit of
Facebook and its users.